\documentclass[twocolumn,showpacs,preprintnumbers,amsmath,amssymb]{revtex4}

\usepackage{graphicx}
\usepackage{dcolumn}
\usepackage{bm}

\begin{document}


\title{\boldmath
Searching for the Bottom Counterparts of $X(3872)$ and $Y(4260)$
via $\pi^+\pi^-\Upsilon$ }
%
\author{Wei-Shu Hou}
\affiliation{Department of Physics, National Taiwan
University, Taipei, Taiwan 10617, R.O.C. }%
%

%
%
\vfill
\begin{abstract}
The $X(3872)$ and $Y(4260)$, among a host of charmonium-like
mesons, have rather unusual properties:
the former has very small total width, the latter has large rate
into $\pi^+\pi^-J/\psi$ channel. It would not be easy to settle
between the many suggested explanations for their composition. We
point out that discovering the bottom counterparts should shed
much light on the issue. The narrow state can be searched for at
the Tevatron via $p\bar p \to \pi^+\pi^-\Upsilon + X$, but the LHC
should be much more promising. The state with large overlap with
$\Upsilon$ can be searched for at B factories via radiative return
$e^+e^- \to \gamma_{\rm ISR} + \pi^+\pi^-\Upsilon$ on
$\Upsilon(5S)$, or by $e^+e^- \to \pi^+\pi^-\Upsilon$ direct scan.
\end{abstract}
\maketitle


 \noindent {\bf 1. Introduction}
 \vskip0.12cm

Owing to the unprecedented luminosities achieved at the B
factories, heavy quarkonium spectroscopy is experiencing a
renaissance. There is an $X$ and a $Y$ and a $Z$ of 3940 MeV
states produced via various mechanisms, but two states stand out
especially: the $X(3872)$ and the $Y(4260)$, both observed in the
$\pi^+\pi^- J/\psi$ channel.

The $X(3872)$ was discovered~\cite{X3872} by the Belle experiment
in $\pi^+\pi^- J/\psi$ recoiling against $K^+$ from $B^+$ decay,
and was quickly confirmed by the CDF~\cite{XCDF} and D0~\cite{XD0}
experiments in $p\bar p \to \pi^+\pi^- J/\psi +$ anything, as well
as by~\cite{XBaBar} the BaBar experiment. The width is narrow,
$\Gamma < 2.3$ MeV~\cite{X3872}, and is consistent with
experimental resolution. Subsequent studies strongly favor the
$1^{++}$ quantum number~\cite{X_1++}. With its mass just at the
$D^0\bar D^{*0}$ threshold, theoretical interpretation has ranged
from $D^0\bar D^{*0}$ molecule, 4 quark state, to charmonium
hybrid.

The $Y(4260)$ was discovered~\cite{Y4260} by the BaBar experiment
in initial state radiation (ISR, or ``radiative return") $e^+e^-
\to \gamma_{\rm ISR} + \pi^+\pi^- J/\psi$ events, hence is
$1^{--}$. The width is found to be around 90 MeV. What is peculiar
is the large partial width for $Y(4260) \to \pi^+\pi^- J/\psi$.
Furthermore, it falls at a local minimum of the $e^+e^- \to$
hadrons cross section. The state has been confirmed~\cite{CLEOc}
by the CLEO-c experiment via $e^+e^- \to \pi^+\pi^- J/\psi$ energy
scan. Theoretical interpretations range from hybrid~\cite{hybrid},
4 quark state~\cite{4quark}, meson molecule~\cite{molecule} or
baryonium~\cite{baryonium}, to conventional
$\psi(4S)$~\cite{psi4S}.

It is not our intention to comment on the various theoretical
interpretations, which clearly needs more data and more debate to
settle. Judging from the history of hadronic spectroscopy, it
would not be easy for this to be conclusive. Rather, the intent of
this short note is to point out where and how to find analogous
states involving $b$ quarks, dubbed the $X_b$ and $Y_b$,
respectively. Since $m_b$ is much larger than $m_c$, observing
such states would not only be spectacular, but should offer
immense help to distinguish between models.

Clearly, the analogous search channel would be $\pi^+\pi^-
\Upsilon$.
%
We point out that the narrow state $X_b$ can be searched for at
the Tevatron (and better at the LHC). The $1^{--}$ state $Y_b$ can
be searched for at the B factories (and future Super B factory),
either by ISR search on the $\Upsilon(5S)$, or by direct scan at
$\Upsilon(5S)$ energies and beyond.

 \vskip0.2cm
 \noindent {\bf 2. {\boldmath $\pi^+\pi^- \Upsilon$ Search at Hadron
 Colliders}
 }
 \vskip0.12cm

Let us first focus on the $X_b$. If this is a $1^{++}$ state, and
unlike the $X(3872)$ case where there is now no analogue of the
parent $B$ meson, one can only think of searching at hadronic
colliders.

The crucial question is: What is the mass?
From the fact that the $X(3872)$ is right at the $D^0\bar D^{*0}$
threshold, the analogy would be the $B\bar B^{*}$ threshold, which
would be at 10604 MeV, regardless of $B^0$ or $B^+$. This is of
course just a guess~\cite{Petrov}. By coupled $s$-- and $d$--wave
$B\bar B^*$ channels, some models predict~\cite{Tornqvist,Swanson}
the $X_b$ mass to be 10562 MeV, below the $B\bar B^*$ threshold,
while the $X(3872)$ mass is brought about by couplings between the
$D\bar D^*$, $\rho J/\psi$ and $\omega J/\psi$
channels~\cite{Swanson}. Whether $M_{X_b} \sim$ 10604 MeV or 10562
MeV, the available energy for the $\pi^+\pi^-$ system is over 1000
MeV, and one can check whether $\rho$ is still dominant once $X_b$
is observed.

Turning to $X_b$ production, we first note that, at the Tevatron,
$J/\psi$ production from $B$ decay is but a fraction of the total
cross section for $p\bar p \to J/\psi +$ anything, while $X(3872)$
production is consistent with $\psi(2S)$ in prompt production
fraction. Therefore, in moving to the $X_b \to \pi^+\pi^-\Upsilon$
search, we assume that the $X_b$ production mechanism is similar
to prompt $X(3872)$ production. For $X(3872) \to \pi^+\pi^-
J/\psi$ reconstruction at CDF~\cite{XCDF}, we will take the number
to be $\sim 3500$ events per fb$^{-1}$.

$\Upsilon$ production at Tevatron energies (for our purpose, we do
not distinguish between 1.8 and 1.96 TeV) has been studied by both
CDF and D0~\cite{UpsCDF,UpsD0}. Compared to $J/\psi$
production~\cite{psiCDF}, the cross section is smaller by almost 3
orders of magnitude. Assuming this fraction, together with the
leptonic rate of $\Upsilon$ being only 40\% that of $J/\psi$, our
very rough estimate for the number of reconstructed $X_b\to
\pi^+\pi^-\Upsilon$ events is of order 20 for an integrated
luminosity of order 8 fb$^{-1}$ expected for the Tevatron Run II.
Thus, the case appears to be marginal.

We caution that we could be off by an order of magnitude, so the
direction should still be pursued at the Tevatron. We do not know,
for example, the branching fraction of $X_b \to
\pi^+\pi^-\Upsilon$ compared with $X(3872) \to \pi^+\pi^- J/\psi$,
nor do we know the variation in production fraction with $m_Q$.
The search program should start with reconstructing $\Upsilon(2S)
\to \pi^+\pi^-\Upsilon$. Making a similar estimate as above,
taking into account the $\Upsilon(2S) \to \pi^+\pi^- \Upsilon$
branching ratio compared with $\psi(2S) \to \pi^+\pi^- J/\psi$,
one expects $\sim 100$ reconstructed $\Upsilon(2S) \to
\pi^+\pi^-\Upsilon$ events for an integrated luminosity of order 8
fb$^{-1}$. If one cannot even establish $\Upsilon(2S)$, then it
would be doubtful whether $X_b$ can be found via the
$\pi^+\pi^-\Upsilon$ channel.

The situation should be much better at the LHC. It is not clear
what is the actual ratio of inclusive $\Upsilon$ vs $J/\psi$
production, although it should be better than at Tevatron
energies. The PYTHIA based simulation results of
Ref.~\cite{Domen}, extrapolating from fitted results to Tevatron
measurements, suggest that the cross section for ${\cal B}
(\Upsilon \to \mu^+\mu^-) \, d\sigma(pp\to \Upsilon + X)/dp_T$ at
LHC is roughly 1/10 that of ${\cal B} (J/\psi \to \mu^+\mu^-) \,
d\sigma(p\bar p\to J/\psi + X)/dp_T$ at the Tevatron. Thus, even
with a few fb$^{-1}$ at the LHC, the ATLAS and CMS experiments
should be able to discover $X_b \to \pi^+\pi^-\Upsilon$, if it
exists and is as narrow as $X(3872)$.
Once again, the benchmark test should be to reconstruct
$\Upsilon(2S) \to \pi^+\pi^-\Upsilon$, and to look for extra
narrow states above it that do not fit the usual $\Upsilon(nS)$
spectrum.

The production of $b\bar b$ is enhanced in the forward direction
at high energy hadronic colliders, and dedicated $B$ experiments
such as LHCb~\cite{LHCb} have a forward detector design aimed at
reconstructing both $b$ hadrons. Thus, LHCb may be the best suited
for the study of $X_b$. It is the only hadronic collider
experiment that has particle identification and full calorimetry
capabilities. Though not needed for $X_b \to \pi^+\pi^-\Upsilon$
search, these should enable it to do a more complete study (such
as $K^+K^-\Upsilon$ or $\omega\Upsilon$) of bottomonium
spectroscopy beyond the $X_b$, such as searching for $d$-wave
mesons which branch into $\pi^+\pi^-\Upsilon$. Once found, the
$J^{PC}$ quantum numbers can be established through, for example,
partial wave analysis. Identifying more states would clearly help
the interpretation.

In preparing for a search for $X_b$ at LHCb, once again the
benchmark test would be to reconstruct $\Upsilon(2S) \to
\pi^+\pi^-\Upsilon$. If LHCb can demonstrate this, given higher
cross section for forward vs central $b\bar b$ production and a
more specialized detector, it should be straightforward to find
the $X_b$, if it exists, while LHCb may be able to discover other
narrow states. It would be interesting to see LHCb shed light on
heavy quarkonium spectroscopy, even though it was designed for
flavor physics.

 \vskip0.2cm
 \noindent {\bf 3. {\boldmath $\pi^+\pi^- \Upsilon$ Search at
 $e^+e^-$ Colliders}
 }
 \vskip0.12cm

The $Y(4260)$ (we shall denote it $Y_c$) was first
observed~\cite{Y4260} in radiative return $e^+e^- \to \gamma_{\rm
ISR}+ \pi^+\pi^-J/\psi$, and confirmed~\cite{CLEOc} by direct
$e^+e^- \to \pi^+\pi^-J/\psi$ scan. The observed width of 88 MeV
is broad compared to $X(3872)$. Averaging over BaBar and CLEO, one
has,
\begin{equation}
\Gamma(Y_c\to ee){\cal B}(Y_c\to \pi^+\pi^-J/\psi) \sim 6\ {\rm
eV},
\end{equation}
or ${\cal B}(Y_c\to ee){\cal B}(Y_c\to \pi^+\pi^-J/\psi) \sim
7\times 10^{-8}$, which is larger than the case for $\psi(4040)$
and $\psi(4160)$. But since $Y_c(4260)$ falls at a dip in the
$e^+e^-\to$ hadrons cross section, barring subtle interference
effects~\cite{psi4S}, presumably $\Gamma(Y_c\to ee) \ll
\Gamma(\psi(4160)\to ee) \sim$ 770 eV. Hence, the partial width
$\Gamma(Y_c\to \pi^+\pi^-J/\psi)$ should be a few MeV or higher,
much larger than typical charmonia.

For $Y_b$, one can contemplate production in radiative return
$e^+e^- \to \gamma_{\rm ISR}\pi^+\pi^-\Upsilon$, or by direct
$e^+e^- \to Y_b\to \pi^+\pi^-\Upsilon$ scan. Search in the
hadronic environment would be hampered by large background due to
a sizable width. The question again is, what is the mass? Further,
what is the width, and $\Gamma(Y_b\to ee){\cal B}(Y_b\to
\pi^+\pi^-\Upsilon)$?

The results of the CLEO study~\cite{CLEOc} of 15 decay modes of
the $Y_c(4260)$ are compatible with the hybrid charmonium
picture~\cite{hybrid}, could be supportive of 4 quark
states~\cite{4quark} if partners are seen, and disfavors all other
proposals. Without advocating a hybrid interpretation, we take the
$Q\bar Qg$ hybrid picture as a guide for discussing the mass of
$Y_b$. Lattice studies have put the lowest $b\bar bg$ hybrid at
around 10700--11000 MeV~\cite{Michael}. The $1^{--}$ quantum
number is possible, with many other possible quantum numbers,
including exotic ones such as $1^{-+}$. The $1^{--}$, however, can
mix with standard $s$-wave mesons and may not be the lightest, but
it is clearly the most accessible.

Lattice studies tend to give lightest $c\bar cg$ hybrid mass
around 4400 MeV. If $Y_c(4260)$ is indeed dominantly a hybrid, by
analogy the $b\bar bg$ hybrid lattice range could be scaled down
to 10600--10900 MeV. This would make $\Upsilon(5S)$, at 10865 MeV,
an excellent place to conduct $e^+e^- \to \gamma_{\rm ISR}Y_b\to
\gamma_{\rm ISR} + \pi^+\pi^-\Upsilon$ search, aside from the main
program of $B_s$ studies. We shall take 10600, 10700 and 10800 MeV
as nominal $M_{Y_b}$ values for this purpose.
We caution, however, that even with lattice studies of hybrids,
there are uncertainties due to difference in numerical approach,
scale uncertainty, as well as treatment of dynamic quarks. For
example, some studies~\cite{Kuti} find the lowest $b\bar bg$
hybrid mass to be $\sim 10900$--11000 MeV, while giving the right
mass for $c\bar cg$ hybrid that is consistent with $Y_c(4260)$. If
$Y_b$ is heavier than 10900 MeV, then a direct scan would be more
profitable.

It is reasonable to assume that $\Gamma(Y_b\to
\pi^+\pi^-\Upsilon)$ is comparable to $\Gamma(Y_c(4260) \to
\pi^+\pi^-J/\psi)$. For the total width, taking $\Gamma_{Y_b} \sim
\Gamma_{Y_c} \sim$ 100 MeV is also reasonable. But $\Gamma_{Y_b}$
could be narrower. For example $Y_c\to D\bar D^*\pi$ is not
forbidden, but $Y_b\to B\bar B^*\pi$ could be hampered by phase
space if $Y_b$ is lighter than $\Upsilon(5S)$. A narrower width
could compensate for the suppression of $\Gamma(Y_b \to ee)$ due
to $b$ quark charge. We therefore take Eq. (1) and estimate that
$\Gamma(Y_b\to ee){\cal B}(Y_b\to \pi^+\pi^-\Upsilon) \lesssim 6$
MeV.

The Belle experiment has performed an engineering run on
$\Upsilon(5S)$ in 2005 with 1.86 fb$^{-1}$ data~\cite{Drutskoy},
and has accumulated over 20 fb$^{-1}$ just before 2006 summer
shutdown. The ISR cross section for $e^+e^- \to \gamma_{\rm
ISR}Y_b\to \gamma_{\rm ISR} + \pi^+\pi^-\Upsilon$ on
$\Upsilon(5S)$ resonance, in the narrow $Y_b$ width approximation
and leading order in $\alpha$, is~\cite{Eidelman}
\begin{eqnarray}
\sigma_{\rm ISR} \simeq 3.6\times 10^7 \, \frac{\Gamma_{ee}{\cal
B}_{ \pi^+\pi^-\Upsilon}}{M_{Y_b}} \,
\frac{1}{x}\left(1-x+\frac{x^2}{2}\right)\; {\rm pb},
\end{eqnarray}
where $x = 1 - M_{Y_b}^2/s$ is the energy fraction carried away by
the ISR photon (usually not observed) in the CM frame. The cross
sections for our representative values of $M_{Y_b} = 10600$, 10700
and 10800 MeV are given in Table~I.

\begin{table}[t]
\caption{Cross section for $e^+e^- \to \gamma_{\rm ISR}\,Y_b \to
\gamma_{\rm ISR}\,\pi^+\pi^-\Upsilon$ on the $\Upsilon(5S)$, and
for direct $e^+e^-\to Y_b \to \pi^+\pi^-\Upsilon$, for $M_{Y_b} =$
10600, 10700 and 10800 MeV, in the narrow width approximation. We
take $\Gamma(Y_b\to ee){\cal B}(Y_b\to \pi^+\pi^-\Upsilon)$ to be
6 eV, comparable to Eq. (1). For higher values of $M_{Y_b}$, ISR
from $\Upsilon(5S)$ ceases to be feasible, but direct scan can
still be done, with only a slight drop in cross section with $s$.
 }
\begin{center}
\begin{ruledtabular}
\begin{tabular}{cccc}
process & 10600 & 10700 & 10800
\\ \hline 
$e^+e^- \to \gamma_{\rm ISR}\, \pi^+\pi^-\Upsilon$ &
 0.4 pb  & 0.6 pb  & 1.6 pb   \\
$e^+e^-\to \pi^+\pi^-\Upsilon$ &
 9.1 pb  & 9.0 pb  & 8.8 pb      \\
\end{tabular}
\end{ruledtabular}
\end{center}
\end{table}

Radiative return cross section is ${\cal O}(\alpha)$ suppressed,
but one might enjoy a longer run on the $\Upsilon(5S)$ for reasons
of $B_s$ physics. One could also gain in $1/E_\gamma$ enhancement
when $Y_b$ is closer to $\Upsilon(5S)$, though the narrow width
approximation may start to be questionable. However, we do not
know the width for $Y_b$, so we just use Table~I as a rough guide.
With 30 fb$^{-1}$ on $\Upsilon(5S)$, assuming $\Gamma(Y_b\to
ee){\cal B}(Y_b\to \pi^+\pi^-\Upsilon)$ is similar Eq. (1), even
for $M_{Y_b}\sim 10600$ MeV one expects close to 600
$\pi^+\pi^-\ell^+\ell^-$ events, where $\ell = e,\;\mu$ and
$m_{\ell\ell}$ reconstructing to $M_\Upsilon$. Thus, even for
$\Gamma(Y_b\to ee){\cal B}(Y_b\to \pi^+\pi^-\Upsilon)$ as low as 1
eV, one can get similar significance for $Y_b$ as the BaBar
discovery of $Y(4260)$, where 125 events were obtained from 211
fb$^{-1}$ data on the $\Upsilon(4S)$. It seems that ISR return on
$\Upsilon(5S)$ would definitely find the corresponding $Y_b$ if it
is lighter in mass.

One could also directly scan for $e^+e^- \to Y_b \to
\pi^+\pi^-\Upsilon$, which would likely be the only option for
$Y_b$ heavier than $\Upsilon(5S)$. The cross section is
\begin{eqnarray}
\sigma_0(s) \simeq \frac{12\pi{\cal B}_{ee}{\cal B}_{
\pi^+\pi^-\Upsilon}}{s} \sim \frac{1027}{M^2_{Y_b}\;({\rm GeV})}\;
{\rm pb} \sim 9 \; {\rm pb},
\end{eqnarray}
where $s = M_{Y_b}^2$, and we have taken ${\cal B}_{ee}{\cal B}_{
\pi^+\pi^-\Upsilon}$ to be the same as for $Y_c(4260)$.
With just 13.2 pb$^{-1}$ on the $Y(4260)$, CLEO was able to
observe~\cite{CLEOc} a clean signal of 37 $\pi^+\pi^-J/\psi \to
\pi^+\pi^-\ell^+\ell^-$ events with little background, measuring
$\sigma_0(e^+e^-\to \pi^+\pi^-J/\psi) \simeq 58$ pb, which is
consistent with Eq. (1). If Eq. (1) holds approximately for $Y_b
\to \pi^+\pi^-\Upsilon$, even though ${\cal B}(\Upsilon \to
\ell\ell) \simeq 0.4\, {\cal B}(J/\psi \to \ell\ell)$, the 30
pb$^{-1}$ per energy scan performed by Belle for $\sqrt{s} =
10825$, 10845, 10865, 10885 and 10905 MeV for the $\Upsilon(5S)$
engineering run~\cite{Drutskoy} can already be very useful. This
is especially so if the scan  can be repeated to cover fully the
10700--11000 MeV range, maybe with 30--50 MeV steps, assuming that
$\Gamma_{Y_b}$ is not drastically different from $\Gamma_{Y_c}$.
But since we do not really know $\Gamma(Y_b\to ee){\cal B}(Y_b\to
\pi^+\pi^-\Upsilon)$ nor $\Gamma_{Y_b}$, discovery may still come
first from radiative return studies.

 \vskip0.2cm
 \noindent {\bf 4. Discussion and Conclusion}
 \vskip0.12cm

For the narrow $X_b$ state, one needs high center of mass energy
to produce the heavy quarkonia of interest, especially for the
case of bottomonia. One also needs to associate the quarkonia with
a $\pi^+\pi^-$ pair to form the exotic meson. As was the case for
$X(3872)$, only the Tevatron was able to confirm the discovery by
the B factories. Note that broad states would be too hard to
establish in a high background environment, even for the Tevatron
and the LHC. Combinatoric background would be much higher for
heavy ion collisions.

We have discussed the mass range of 10560--10600 MeV motivated by
$B\bar B^*$ threshold. But for the actual hadronic collider
search, one should certainly aim for a broader range. Since the
case is marginal  at the Tevatron, discovery may have to wait for
the LHC.
Whether Tevatron or LHC, $\Upsilon(2S)\to \pi^+\pi^-\Upsilon$
reconstruction should be studied. We remark that many new states,
such as $d$-wave mesons, are not quite accessible at $e^+e^-$
machines because of suppressed $e^+e^-$ widths. Detectors at the
LHC, especially the forward design of LHCb, have good potential
for discovering other narrow bottomonia beyond the $X_b$ via
$\pi^+\pi^-\Upsilon$ (and charmonia via $\pi^+\pi^-J/\psi$).

More immediately accessible at the B factories is the broad $Y_b$
state that decays prominently into $\pi^+\pi^-\Upsilon$. We have
used the 10600-10800 MeV range motivated by hybrid $Q\bar Qg$
picture to illustrate the efficacy of radiative return or direct
scan search, in part because of the available $\Upsilon(5S)$ data
at Belle. But again, the target range should be broader, as
$M_{Y_b}$ could be in 10900--11000 MeV range. Furthermore, model
pictures such as 4-quark states should also be kept in mind, and
$Y_b$ mass above 11000 MeV is not impossible. But it would be
difficult to persuade B factories to run at such energies.

The best case would be if $Y_b$ is below the $\Upsilon(5S)$.
Unless $\Gamma(Y_b\to ee){\cal B}(Y_b\to \pi^+\pi^-\Upsilon)$ is
much less than 6~eV (Eq.~(1)), the state is likely to be
discovered in a 30 fb$^{-1}$ or so data run. Knowing the mass and
width, one can then do the direct $e^+e^- \to \pi^+\pi^-\Upsilon$
scan and search for other channels such as $K^+K^-\Upsilon$ or
$\pi^0\pi^0\Upsilon$. If $Y_b$ does not show up in radiative
return on $\Upsilon(5S)$, besides the possibility of suppressed
$\Gamma(Y_b\to ee){\cal B}(Y_b\to \pi^+\pi^-\Upsilon)$, it is
possible that $Y_b$ is heavier. A quick scan with 30 pb$^{-1}$
each for the energies 10940, 10980, 11020, 11060, 11100 MeV could
complement the existing scan around $\Upsilon(5S)$, i.e. 10825,
10845, 10865, 10885 and 10905 MeV, and could extend the discovery
potential.
There is a good chance that the $Y_b$ could be discovered soon.

\vskip 0.3cm \noindent{\bf Acknowledgement}.\ This work is
supported in part by NSC 94-2112-M-002-035.

\end{document}